# Strain distribution in WS$_2$ monolayers detected through Polarization-resolved Second Harmonic Generation

**George Kourmoulakis**[1,2], **Sotiris Psilodimitrakopoulos**[1]*, **George Miltos Maragkakis**[1,3], **Leonidas Mouchliadis**[1], **Antonios Michail**[4,5], **Joseph A Christodoulides**[6], **Manoj Tripathi**[7], **Alan B Dalton**[7], **John Parthenios**[5], **Konstantinos Papagelis**[5,8], **Emmanuel Stratakis**[1,3]*, and **George Kioseoglou**[1,2]*

[1] *Institute of Electronic Structure and Laser, Foundation for Research and Technology - Hellas, Heraklion, 71110, Crete, Greece*
[2] *Department of Materials Science and Technology, University of Crete, Heraklion, 70013 Crete, Greece*
[3] *Department of Physics, University of Crete, Heraklion Crete 70013, Greece*
[4] *Department of Physics, University of Patras, Patras, 26504, Greece*
[5] *FORTH/ICE-HT, Stadiou str Platani, Patras 26504 Greece*
[6] *Naval Research Laboratory, 4555 Overlook Ave SW, Washington, DC 20375-5320, U.S.A*
[7] *Department of Physics and Astronomy, University of Sussex, Brighton BN1 9RH, United Kingdom*
[8] *School of Physics, Department of Solid-State Physics, Aristotle University of Thessaloniki, Thessaloniki, 54124, Greece*

* Authors to whom any correspondence should be addressed.

Email: sopsilo@iesl.forth.gr, stratak@iesl.forth.gr, and gnk@materials.uoc.gr

## Abstract

Two-dimensional (2D) graphene and graphene-related materials (GRMs) show great promise for future electronic devices. Nevertheless, GRMs result distinct properties under the influence of the substrate that serves as support through uneven compression/ elongation of GRMs surface atoms. Strain in GRM monolayers is the most common feature that alters the interatomic distances, band structure, providing a new degree of freedom that allows regulation of their electronic properties and introducing the field of straintronics. Having an all-optical detection, a minimally invasive tool that rapidly probes strain in large areas of GRM monolayers, would be of great importance in the research and development of novel 2D devices. Here, we use Polarization-resolved Second Harmonic Generation (P-SHG) optical imaging to identify strain distribution, induced in a single layer of WS$_2$ placed on a pre-patterned Si/SiO$_2$ substrate with cylindrical wells. By fitting the P-SHG data pixel-by-pixel, we produce spatially resolved images of the crystal armchair direction. In regions where the WS$_2$ monolayer conforms to the pattern topography, a distinct cross-shaped pattern is evident in the armchair image owing to strain. The presence of strain in these regions is independently confirmed using a combination of atomic force microscopy and Raman mapping.

**Introduction**

Monolayers of Transition Metal Dichalcogenides (TMDs) are semiconducting materials with exciting optical and electronic properties[1–3]. Single layer TMDs have a direct band gap meaning that by providing the appropriate excitation energy, strong light emission, through radiative relaxation, can be achieved even at room temperature[4,5]. In addition, the lack of inversion center in atomically thin TMDs leads to large second-order susceptibility ($\chi^{(2)}$) [6–9], making them ideal for studies in the non-linear regime. Second Harmonic Generation (SHG) is the lowest frequency non-linear optical process in which two photons are converted into one with double the frequency of the incident photons. The SHG signals result from the non-centrosymmetric nature of the armchair direction of the crystal[10–12]. The resulting SHG signals are coherent with well-defined polarizations. Through polarization analysis (Polarization-resolved Second Harmonic Generation, P-SHG), SHG microscopy provides the armchair orientation[13]. This procedure, combined with an imaging technique, can provide essential information about the crystal structure of non-centrosymmetric 2D materials, paving the way for the development of advanced non-linear optoelectronic applications[14]. Indeed, P-SHG imaging has been recently demonstrated as an all-optical tool that can provide quantitative information on the 2D crystal quality[13,15], valley population imbalance[16], as well as the relative (twist) angle between overlapping monolayers[17,18].

SHG microscopy is used to produce images of non-centrosymmetric structures, such as 2D TMD crystals. Raman and PL microscopy are spectroscopic imaging techniques that use the Raman scattering and photoluminescence (PL) processes to provide information about molecular vibrations and electronic transitions, respectively. The main advantages of SHG microscopy over Raman and PL microscopy are the following: SHG imaging is a rapid characterization technique, with image acquisition time reported to be around four orders of magnitude smaller than that for Raman mapping[19]. SHG microscopy has high axial and lateral resolution, as it is a coherent process that depends on the square of the electric field intensity, unlike Raman and PL microscopy, which are incoherent processes that depend on the linear intensity. SHG microscopy has high specificity and contrast, as it only occurs in non-centrosymmetric structures, unlike Raman and PL microscopy, which can occur in any structure that has molecular or electronic transitions[20]. Additionally, P-SHG applied pixel-by-pixel, in pixels with size much smaller than the beam waist radius (dictating the optical resolution), provides super-resolved mappings of the SHG orientational fields, enabling a form of optical nanoscopic imaging (probing the crystal orientation of the materials)[21].

The membrane nature of TMDs gives rise to several possibilities for tailoring their properties. Broken inversion symmetry can be effectively tuned following chemical, mechanical and/or optical approaches[22–24]. Several reports focus on the enhancement of the SHG signal from monolayer TMDs. For example, some have focused on the excitonic energy of the material following on-resonance studies[25,26], while others have embedded membranes into specifically designed micro and nanoscale cavities achieving thousand-fold enhancement owing to cavity modes[27,28]. Another popular physical approach to tune SHG signal is to reduce in-plane symmetry with strain in 2D materials[29]. Strain can directly affect the crystal structure by altering bond lengths and angles, tailoring the band gap and lattice constants which leads to modification of their optoelectronic properties[30–35]. Moreover, 2D materials have the ability to tolerate larger magnitudes of mechanical strain than their bulk counterparts[33,36]. In addition, their flat form qualifies them for controlled strain studies (e.g. uniaxial, biaxial, tensile or compressive) by transferring them on dedicated devices and/or patterned substrates for a pre-fixed form and value of strain[34,35,37].

Several experimental techniques can be utilized to identify and study strain in these materials for which fundamental studies are of high importance. For example, in Raman spectroscopy of semiconducting TMDs, the $E_{2g}$ vibrational mode is more sensitive to strain than $A_{1g}$[38–40]. As a result, by applying tensile (compressive) strain, the frequencies of the prominent vibrational modes ($E_{2g}$, $A_{1g}$) decrease (increase). Furthermore, for anisotropic strains (e.g. uniaxial), the degeneracy of the $E_{2g}$ mode can be lifted[41]. Additionally, in photoluminescence (PL) spectroscopy, direct to indirect bandgap transitions[42] and interplay between the intensity emission of neutral, charged and indirect excitons in monolayer TMDs has been reported as a result of strain application[43]. Moreover, SHG has been recently used to probe strain in TMDs. In particular, the polar pattern of P-SHG is elongated along the strain axis, offering a fingerprint of strain onto the P-SHG properties [44–47]. Therefore, P-SHG imaging can be used for identifying and characterizing strain induced in 2D TMDs by their placement on a substrate in a complementary manner to the strain characterization techniques.

In this work, we investigate the SHG signals of mechanically exfoliated $WS_2$ monolayers placed over pre-patterned $Si/SiO_2$ substrates with 3μm diameter cylindrical wells. During the sample fabrication process, the presence or absence of trapped air in the wells creates suspended or conformed (to the well topography) monolayer regions, respectively. Using AFM imaging and Raman mapping, it was found that strain is induced in the monolayer areas conformed firmly into the wells. In these suspended and conformed $WS_2$ monolayer areas, an all-optical P-SHG imaging was performed. Then the P-SHG data were fitted pixel-by-pixel (for every point of the sample) into a theoretical model that calculates the crystal armchair direction of the monolayer. This methodology reveals a characteristic cross-shaped pattern as the fingerprint of non-uniform strain in the armchair images of the $WS_2$ monolayer firmly conformed regions.

It is, therefore, demonstrated that P-SHG imaging is a powerful optical tool for the characterization of strained monolayer TMDs over periodically patterned substrates by providing their fingerprint in the P-SHG contrast. Since strain is commonly used to tune and accomplish desired electronic properties in these 2D materials, P-SHG imaging can become valuable for developing straintronics in advanced multifunctional devices.

**Results and discussion**

$WS_2$ monolayer is transferred on top of a pre-patterned $SiO_2$ substrate with 3μm diameter cylindrical wells (see Materials and Methods). In the Supplementary Material, a typical Raman spectrum is presented, as recorded using an excitation wavelength of 473 nm (Supplementary Fig. S1) that verifies the presence of a $WS_2$ monolayer. In Fig. 1a, we present an optical microscope image of the sample, where the region of the wells covered with the $WS_2$ monolayer is circumscribed by the white dashed line in a polygon shape. The light blue and yellow regions on the top right of the image correspond to bulk material. We observe that many of the covered wells appear optically dark, while the rest are brighter than the uncovered ones at the bottom of the image. The difference in the optical contrast here is resulting from light interference effects[2, 48, 49, 50–52]. The AFM height profile along two adjacent wells within the covered region shown in Fig. 1b, reveals areas where the monolayer membrane is suspended (optically dark well) or fully conformed (optically bright well). In the optically dark contrasted wells, the monolayer rests 30nm from the top surface. In optically bright wells, it is fully conformed to the depth (~174 nm) and the periphery of the well (inset of Fig. 1b). Therefore, it can be concluded that during the transfer of the monolayer onto the pre-patterned substrate, the air is trapped inside most of the cylindrical wells. Trapped air can suspend the monolayer membrane or, in some wells, the absence (or loss of air) of air-induced pressure leads the monolayer to adhere at the well bottom and partially at its periphery resulting in a firmly conformed membrane. The sinking of $WS_2$ layers towards the inside walls of the wells could be responsible for its straining, which will be characterized through Raman spectroscopic mapping in the following section.

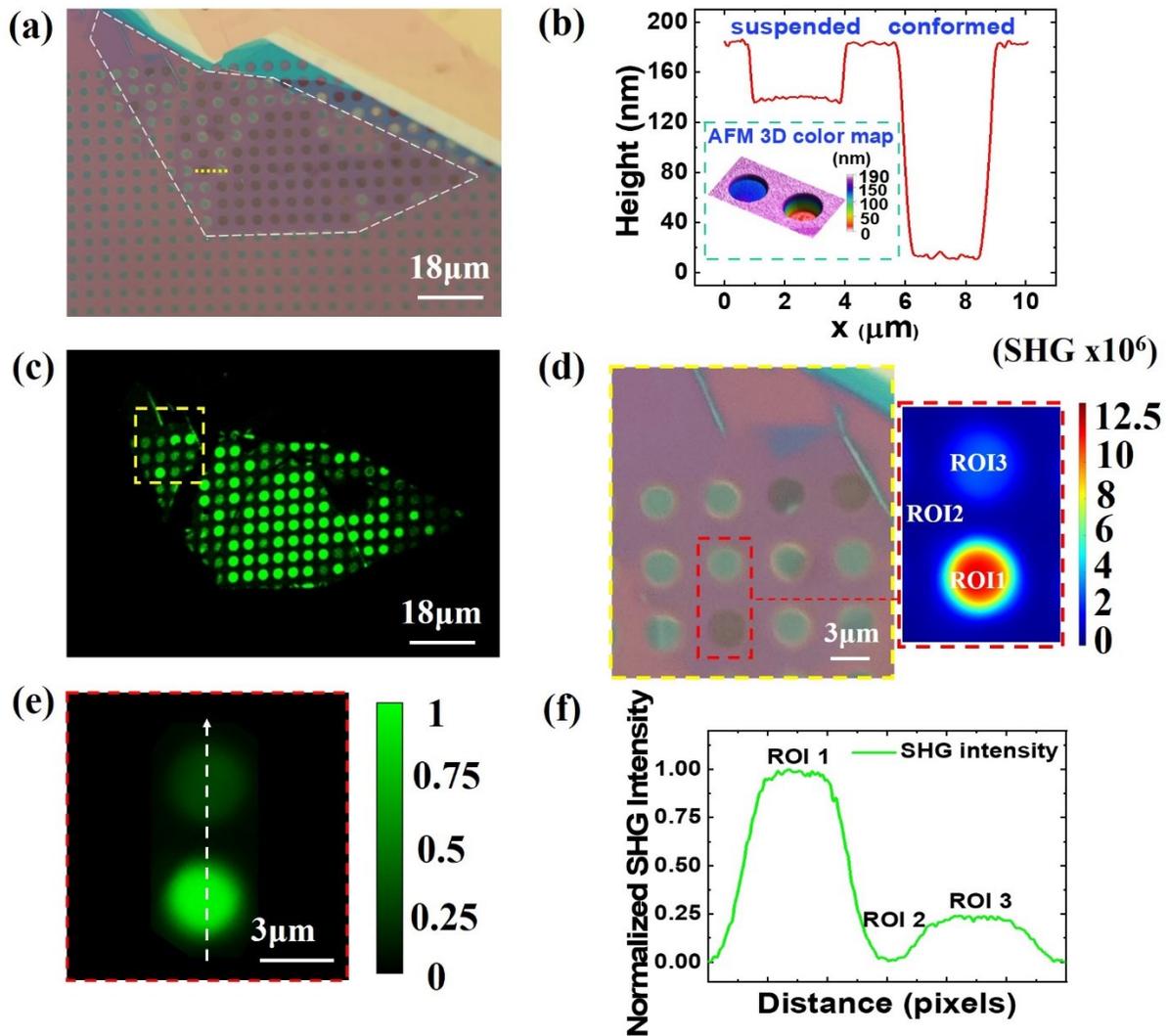

**Figure 1**. Comparison between wide field and SHG imaging for suspended and unsuspended $WS_2$ monolayer regions. (a) Optical image of monolayer $WS_2$ transferred on top of cylindrical wells (monolayer is denoted by white dashed line). (b) AFM height profile along the yellow dashed line in (a) that connects the diameters of two adjacent wells where the monolayer is suspended (optically dark well) or conformed (optically bright well) (Inset: AFM 3D topography of these two wells). (c) SHG intensity mapping of the sample (yellow dashed rectangle denotes a region of interest with several optically bright and dark wells). (d) Optical image of the region denoted in (c). SHG intensity color map is presented for the three ROIs: ROI1-suspended, ROI2-supported outside the well, and ROI3-conformed/non-uniformly strained. (e) Selection of a line scan in SHG (normalized) signal mapping. (f) SHG intensity profile plot of ROIs. It should be highlighted that the SHG signals exhibit approximately three times greater strength in wells containing a suspended monolayer compared to those with a conformed/non-uniformly strained layer.

The source of SHG in monolayer $WS_2$ is the non-centrosymmetric spatial distribution of the W, S atoms in the crystal armchair direction. Although the SHG signals are expected to be generated in the forward propagation direction, for a single non-linear dipole (like in atomically thin WS$_2$ monolayers), equal generated forward- and backwards-SHG signals are expected[53]. We have verified the above by placing WS2 monolayer in thin sapphire substrate and recording almost equal intensity SHG signals in forward and backwards propagation directions (data not shown).

In order to characterize these regions on top of the cylindrical wells with spectroscopy means, we perform PL measurements that are commonly used to investigate suspended monolayers[54]. The PL measurements are performed at room temperature using a CW 543 nm excitation wavelength. We find that the PL intensity in the well with suspended WS$_2$ monolayer is ~tenfold enhanced with respect to the one originating from wells of the conformed monolayer (Supplementary Fig. S2).

The comparative study based on the SHG intensity from suspended, conformed and supported (on un-patterned Si/SiO$_2$) areas of the WS$_2$ monolayer is performed to identify the effect of the patterned substrate. We observe two differently contrasted areas in the SHG image of the monolayer on top of the cylindrical wells (Fig. 1c), namely bright and darker regions. By direct comparison of Fig 1 (a) and (c), it is readily observed that the bright areas in the SHG image (Fig. 1c) correspond to the optically dark areas in the optical image (Fig. 1a), while the darker areas in the SHG image correspond to the optically bright areas in the optical image. Therefore, we conclude that the suspended regions of the WS$_2$ monolayer produce stronger SHG signals than the conformed ones.

We choose three regions of interest (ROIs) of WS$_2$ monolayer to study, which are contained in the red dashed line frame in Fig. 1d. ROI 1 corresponds to the suspended region, ROI 2 corresponds to the region supported by the substrate outside the wells, and ROI 3 corresponds to the part that conforms to the well topography.

The SHG intensity contrast image (Fig. 1e) and its corresponding profile (Fig. 1f) taken along the dashed line in Fig. 1e, illustrate a significant distinction between suspended, conformed, and supported WS$_2$ (i.e., Si/SiO$_2$ regions between two wells). The intensity over the suspended region is significantly high; a ~14-fold enhanced with respect to the supported one and a ~3-fold enhanced compared to the conformed (strained region). This finding is remarkable considering that the enhanced SHG intensity is neither plasmon-mediated nor resulting from cavity mode mediation in contrast to other reported studies on pre-patterned substrates [26,27]. The shape and the uniformity of dielectric environment can have an impact on the absorption and emission properties of 2D materials [48,49]. In our case the distance of the monolayer WS$_2$ from the Si/SiO$_2$ interface varies in suspended and conformed regions. The bottom of the well will act as a mirror and owing to interference effects we observe these topological differences regarding the SHG intensity signal. Therefore, by suspending a 2D TMD (in specific distances from the substrate) and considering each time the excitation wavelength and the refractive indices involved, it is possible to tune its non-linear optical response. In this context, our experimental results are substantial evidence of the underlying substrate role in the non-linear optical properties of 2D TMD systems.

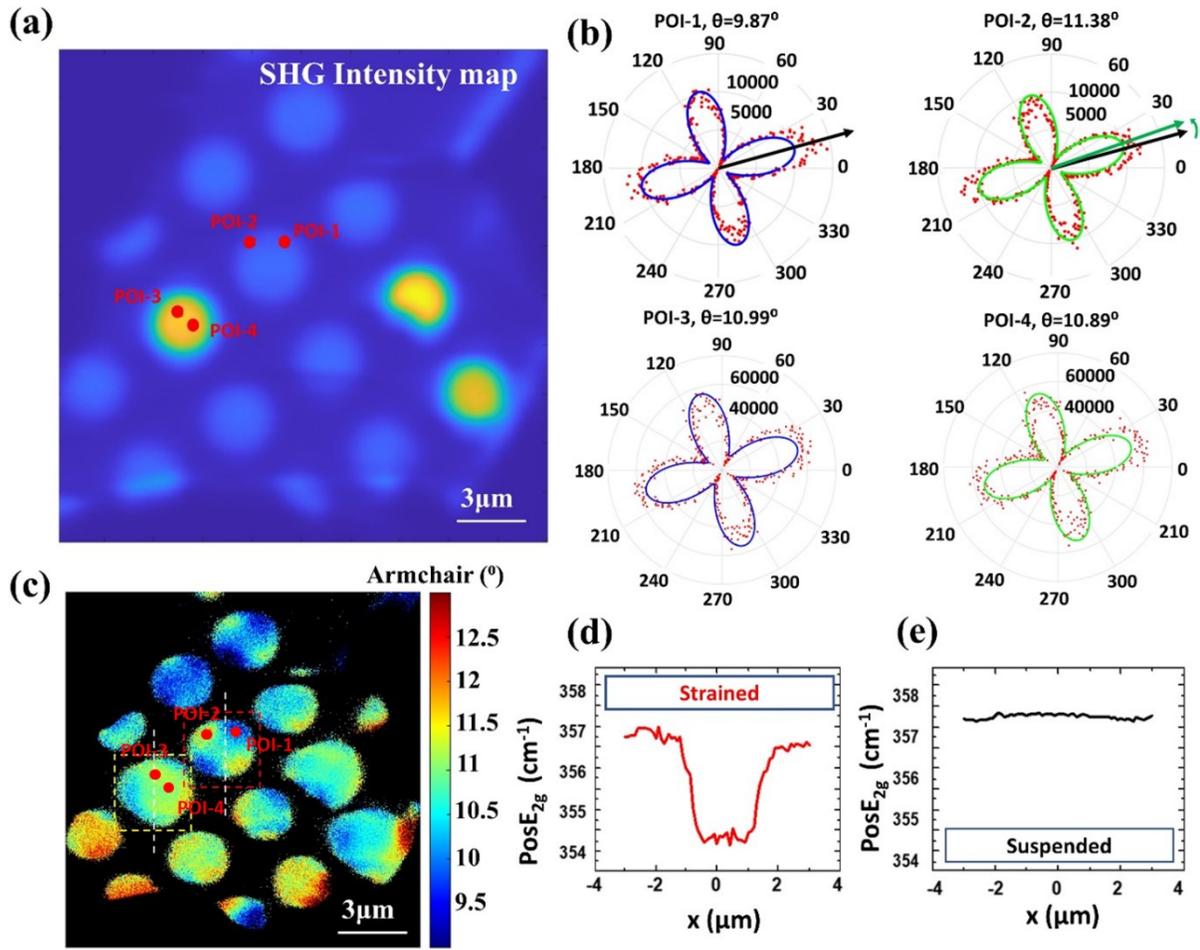

**Figure 2**. Comparative study focusing on suspended and conforming areas. (a) SHG intensity mapping of an area containing 4 pixels of interest (POIs) denoted by red dots. POI-1 and POI-2 are both located at the well where the monolayer is conformed while both POI-3 and POI-4 are at the well where the monolayer is suspended. (b) Comparison of the corresponding polar plots of the POIs presented in panel (a). We note that by comparing the polar plots POI-1 and POI-2, which belong in the same well-1, there is a significant shift in the armchair orientation of the 2D crystal between those POIs (i.e. POI-1 $\theta_1=9.87°±0.18$, $R^2=0.89$ and POI-2 $\theta_2=11.38°±0.18$, $R^2=0.90$), while by comparing POI-3 and POI-4, belonging in the same well-2 we did not observe a significant shift in the armchair orientation of the 2D crystal between those POIs (i.e. POI-3 $\theta_3=10.99°±0.18$, $R^2=0.85$ and POI-4 $\theta_4=10.89°±0.18$, $R^2=0.84$). (c) Image of the armchair orientation pixel-by-pixel mapping for the same area depicted in (a). Red and yellow dashed lines mark the regions of interest. A noticeable difference in the armchair distribution is presented by comparing conformed (i.e. well-1) and suspended areas (i.e. well-2). We clearly observe a cross pattern in well-1 under strain. (d-e) Corresponding Raman mappings of $E_{2g}$ mode for the marked wells in (c). Raman measurement confirmed that the P-SHG cross pattern appears in the well-1 with a strained monolayer.

In our further investigation, P-SHG microscopy is implemented to image pixel-by-pixel the crystal's armchair orientation using the methodology described in previous work [13]. We focus on two wells of different SHG intensity signals to investigate the degree of uniformity regarding the armchair distribution, see Fig. 2.

It is carried out by monitoring the specific field of view where suspended, and strained monolayer $WS_2$ regions are in proximity. Our experimental setup allows for real-time data acquisition and a large field of view, making it easier to identify regions where comparison can be straightforward. We show the SHG intensity map in Fig. 2a, where we mark 4 pixels of interest (POIs). The first two (POI-1 and POI-2) are located in well-1, where the monolayer is conformed to the shape. The other two points (POI-3 and POI-4) are located in well-2, where the monolayer is suspended. The selection of these regions is related to the investigation of the armchair orientation and, more specifically, the degree of its uniformity utilizing P-SHG. In Fig. 2b, we present the polar plots of the POIs 1-4, where the red dots correspond to experimental data while the solid line is the fitting (see Materials and Methods) of the theoretical model that describes the P-SHG modulation from $WS_2$ monolayers[9]. The optical resolution of the P-SHG setup is dictated by the numerical aperture (NA) of the objective lens (1.3 NA) and the excitation wavelength (1030nm), resulting in limit of resolution, L.R. = 483nm. We used the Rayleigh criterion for the limit of resolution, L.R. = 0.61λ/NA, where λ is the excitation wavelength (1030nm) and NA (1.3) is the numerical aperture of the focusing objective.

In Fig. 2b, we observe an angle difference at the order of ~1.5° between POI-1 and POI-2 (both points are located at the conformed area of well-1). Here we only consider the rotation of the polar diagram which corresponds to the armchair orientation, and we do not consider changes in the shape of the polar plot. This rotation of the armchair orientation is related to mechanical strain [29]. The accuracy of the P-SHG technique in the determination of the armchair orientation is the instrument uncertainty and has been calculated as 0.16°[13]. In the suspended area, the polars (from POI-3 and POI-4) do not reveal any difference in orientation due to the expected absence of strain in well-2. Every pixel in the image has its own polar plot. By fitting those polars in a pixel-by-pixel manner for all the pixels of the image and by keeping only the pixels which polars presented a quality of fitting ($R^2$) bigger than 0.36, we produce a spatially resolved image of the armchair orientation (θ) for wells 1-2, under study (Fig. 2c). It is important to note here that the significant variations of the armchair images observed in Fig 2c around the areas of focus marked with dashed squares, are due to cracks/slits of the 2D material that occurred during the transfer process.

Differences in color in the pixels of the armchair directions image (Fig. 2(c)) correspond to different orientations in the P-SHG polar diagrams. This is the origin of the cross-shaped contrast in the strained $WS_2$ regions. To demonstrate our method, we have chosen two pixels (POI-1 and POI-2) belonging in the region where the sample is strained and two pixels (POI-3 and POI-4) from the suspended region. We note that the armchair direction θ for POI-1 is 9.87° while for POI-2 is 11.38°ᵒ. This change in the armchair direction values is due to the shift in the polar orientation, noted in Fig. 2(b) with a green arrow. The changes in the orientations of the polar diagrams for the pixels in the strained region correspond to different armchair direction values, which create the cross-shaped pattern in the colormap. On the other hand, the polar diagrams POI-3 and POI-4 that belong to suspended region, do not exhibit orientation shifts. This is demonstrated by the polar diagrams of Fig. 2(b) POI-3 and POI-4 which result in armchair directions of 10.99º and 10.89º, respectively. These small changes in the spatial distribution of armchair directions provide a uniform contrast for the suspended regions in image Fig. 2(c), unlike the strained ones where the characteristic non uniform cross-shaped pattern appears.

We observe that the armchair direction distributes uniformly in the fully suspended monolayer $WS_2$ above well-2 (see Fig. 2c), with some deviation observed at the rim of the well. On the contrary, the resulting armchair direction image from the conformed monolayer $WS_2$, under non-uniform strain in well-1 (Fig. 2c), exhibits an irregular distribution of the armchair orientation, forming a cross-shaped pattern. This shows that strain can affect the crystal's armchair orientation by probably locally changing the lengths and angles of the atomic bonds[30,32,42]. Importantly, this non-uniform distribution is revealed to exhibit a cross-shaped pattern, which is present systematically over almost all similar flakes under strain. We attribute this to the collapse of the monolayer inside the well and the formation of mechanical strain. In this context, this cross-shaped pattern revealed by P-SHG imaging of armchair orientation ($\theta$) appears to be the signature of non-uniform strain in TMD monolayers conformed to the cylindrical wells.

Experimental errors, e.g., due to possible ellipticity in the excitation polarization (introduced e.g., by the dichroic mirror just before the objective) or the introduction of axial field components because of tight focusing (NA=1.3), are affecting the shape of the experimental P-SHG polar diagrams (different size lobes) or the minimum SHG values in the polar plots (the lobes do not reach zero), respectively. Nevertheless, although the shape of the polar diagrams could be affected by experimental errors, its orientation which is used for extracting the armchair direction, is not expected to be affected. Additionally, experimental errors are systematic among all pixels, hence their effect in the polar diagrams will also be systematic and consequently does not affect the contrast of the armchair direction image.

Different pixel sizes were used in our SHG images. Nevertheless, in all images we are oversampling the 483nm lateral resolution, like in Ref.21, using pixel sizes much smaller than the limit of resolution. This allows for super-resolved orientational fields optical nanoscopy[21].

Raman spectroscopy is used to investigate the presence of mechanical strain in wells 1 and 2. Detailed line scans were conducted across the diameter of the wells with a step of 100 nm. A redshift ($\approx$2.5 cm$^{-1}$) of the $E_{2g}$ mode frequency for the conformed membranes was observed (Fig. 2d). Indeed, considering that the strain sensitivity of the $E_{2g}$ mode is about 6 cm$^{-1}$/%[39], we determine that strain increases abruptly, within 500 nm from the rim towards the well center, up to an almost constant maximum value of 0.4% at the bottom of the well (Fig. 2d). In contrast there is no shift for the suspended membranes (Fig. 2e). The P-SHG cross-shaped pattern appears only in the strained/conformed monolayer.

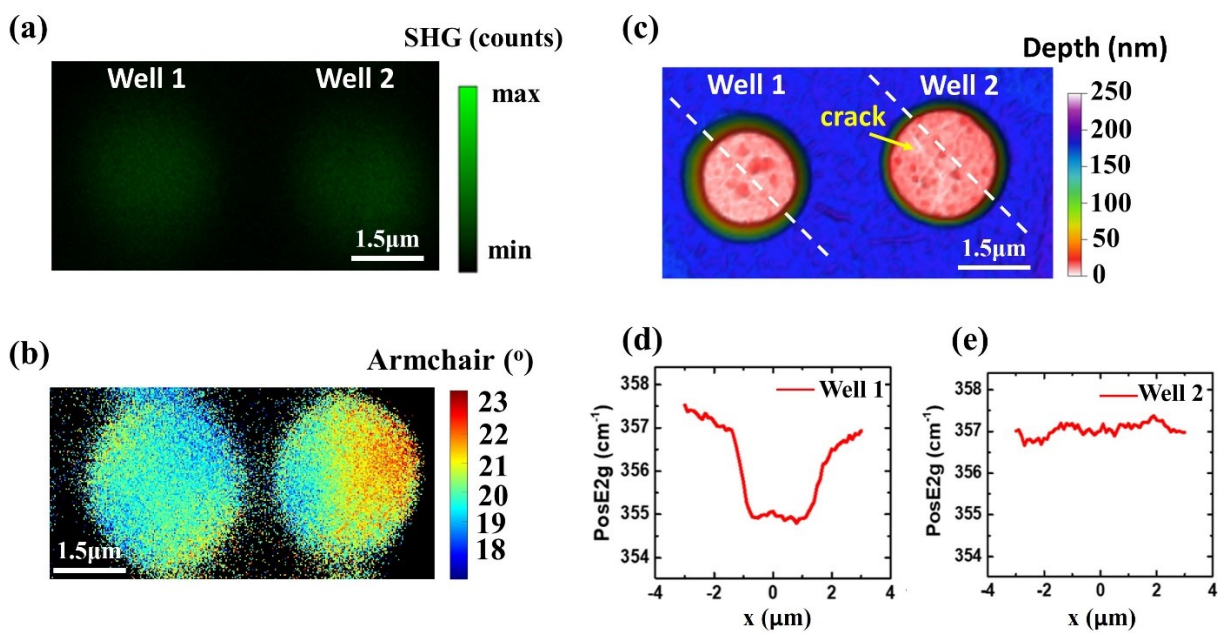

**Figure 3**. Justification that point-by-point mapping of the armchair orientation provides a fingerprint of strain. (a) SHG intensity image of two optically identical areas conformed in the wells 1-2. (b) Armchair orientation mapping from the wells 1-2 seen in Fig. 3a. We note that in well-1, the characteristic cross-shaped pattern indicating strain, is formed. (c) AFM 3D depth color map of well-1 and well-2 respectively. The yellow arrow indicates the crack in well-2 that released the strain in this area and the white dashed lines indicate the line scans for Raman mapping. (d-e) Raman mapping of $E_{2g}$ mode across a line scan (white dashed lines in Fig. 3c) revealing the strain effect for the selected areas. Raman mapping agrees with P-SHG findings that strain is present only in well-1.

An interesting case that provides further justification that the characteristic pattern observed in the P-SHG armchair image is probed by strain, is presented in Fig. 3. While studying several regions, we found a number of optically bright areas (conformed $WS_2$), in which a crack/slit could have released the applied strain (verified by Raman). Based on this finding, we aimed to confirm that the cross-shaped pattern is only present in conformed areas where the strain is preserved. In this context, we selected two optically (bright) similar wells, shown as well-1 and well-2 in Fig. 3a. The SHG signal acquired by both wells was found to be similar, with lower intensity compared to suspended wells, indicating possible strained monolayer areas. However, the P-SHG image in Fig. 3b shows that the cross-shaped pattern associated with strain is present only in well-1. Conducted Raman line scans (indicated by the dashed lines in Fig. 3c), revealed the presence of strain in only well-1 (see Fig 3. d, e). Interestingly, an AFM 3D depth mapping, presented in Fig. 3c, shows that there is a crack in well-2 (also visible and noted by the yellow arrow in Fig. 3c) that most likely released the strain. Remarkably, Raman mapping (Fig. 3d and 3e) proved that only well-1 is under strain, since the $E_{2g}$ in-plane vibrational mode shows a significant variation of 2 cm$^{-1}$ across the well (Fig. 3d). This result validates that point-by-point mapping of the 2D crystal armchair orientation using P-SHG imaging can unambiguously identify strain in a monolayer TMD system.

**Conclusions**

In summary, P-SHG imaging is presented here as a non-invasive all-optical tool that can identify strained 2D monolayers placed above cylindrical wells. We studied the effect of substrate on SHG, by comparing suspended with conformed to the well topology $WS_2$ monolayers. The ~14-fold and ~3-fold enhanced SHG intensity of suspended areas compared to supported and conformed areas respectively was a result of thin film interference effects. Pixel-by-pixel mapping of the crystal armchair orientation based on P-SHG optical measurements, revealed a cross-shaped pattern that is characteristic fingerprint of non-uniform strain in monolayer $WS_2$ over 3μm cylindrical wells. This was independently confirmed by Raman mapping. P-SHG and Raman analysis over (optically bright) similarly conformed areas with and without strain verified our results. Having an all-optical tool able to identify strain in large areas of such systems could be of significant value towards their fast characterization. Our work can potentially be applied to systems with different types of strain and 2D TMD monolayers, becoming a complementary technique towards the design, characterization, and quality control during fabrication of advanced optoelectronic devices.

**Materials and methods**

*A. Sample Preparation*

Polydimethylsiloxane films (PDMS) were fabricated from 10:1 mixing ratio of SYLGARD 182 Silicone Elastomer Kit with heat cure at 80° C for 2 hours. High quality $WS_2$ bulk crystals were purchased from HQ Graphene and mechanically exfoliated directly on the PDMS films. The films were placed on typical microscope glass slides using a standard method. $WS_2$ monolayers were realized under an optical microscope and characterized with Raman spectroscopy. The glass slide with $WS_2$ monolayer was mounted on a XYZ micromechanical stage under a custom coaxially illuminated microscope and transferred on a pre-patterned $Si/SiO_2$ (285 nm) substrate using viscoelastic stamping[55]. The selected pattern consisted of a matrix of cylindrical wells of 3 μm in diameter and 174 nm in depth. They were fabricated following a method based on e-beam lithography.

*B. Experimental Setup*

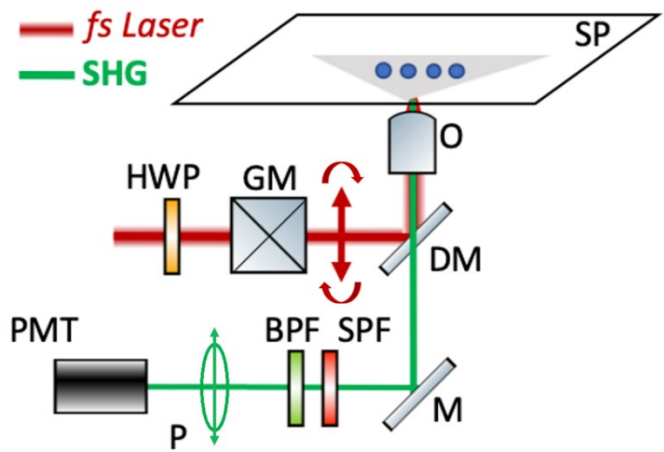

**Figure 4**. Schematic representation of the experimental setup for P-SHG imaging. Abbreviations: HWP: half-wave plate, GM: galvanometric mirrors, DM: dichroic mirror, O: objective lens, SP: sample plane, M: mirror, SPF: short-pass filter, BPF: bandpass filter, P: linear polarizer, PMT: photomultiplier tube.

In Fig.4, our custom-built SHG experimental apparatus is based on a diode pumped Yb:KGW fs oscillator (1030 nm, 30 fs, 76 MHz, Flint, Light Conversion, Vilnius, Lithuania) guided into an Axio Observer Z1 (Carl Zeiss, Jena, Germany) inverted microscope. The laser beam passes through a zero-order half-wave retardation plate (QWPO-1030-10-2, CVI Laser), placed on a motorized rotation stage (M-060.DG, Physik Instrumente Karlsruhe, Germany) that can rotate with high accuracy (0.1°) the orientation of the excitation linear polarization. Raster-scanning of the beam at the sample plane is performed using a pair of silver-coated galvanometric mirrors (6215H, Cambridge Technology, Bedford, MA, USA). The beam is reflected on a short-pass dichroic mirror at 45º (DMSP805R, Thorlabs: Newton NJ, USA) placed at the motorized turret box of the microscope just below the objective (Plan-APO 40x/1.3 NA, Carl Zeiss). The SHG signals are detected in the backwards direction, after passing through a short-pass filter (FF01-680/SP, Semrock, Rochester, NY, USA) and a narrow (3nm) band-pass filter (FF01-514/3, Semrock). A fixed linear polarizer (LPVIS100-MP, Thorlabs) is placed in front of a photomultiplier tube module (H9305-04, Hamamatsu, Hamamatsu city, Japan) to select the polarization state of the detected SHG signals.

Raman spectra were collected with a Renishaw inVia Raman spectrometer in the backscattering geometry. The beam of a solid state 515 nm (Cobolt Fandago) laser was focused by means of a 50× objective lens (N.A. = 0.75). Laser power was kept below 0.1 mW to avoid laser heating effects and photodoping. The Raman scattered radiation was dispersed by a 2400 grooves/mm diffraction grating. Our system uses a Renishaw MS100 encoded motorized XYZ sample stage, allowing collection of Raman maps with a step size of 100 nm.

## C. Fabrication of cylindrical wells on Si/SiO$_2$

A film of 285nm SiO$_2$ was deposited using a plasma-enhanced pressure chemical vapor deposition (PECVD) on a p-doped Si wafer. Arrays of 3-micron diameter disks with a period ranging from 3 to 8 microns were patterned and then etched via an inductively coupled plasma reactive ion etching (ICP-RIE) for a given time to form holes into the SiO$_2$ film, with a depth of 174 nm.

## D. P-SHG theoretical model and fitting procedure

In order to describe the interaction of the excitation laser field with the monolayer WS$_2$ and the production of SHG, we employ the Jones formalism [13,15]. Two coordinate systems are considered: the laboratory (X, Y, Z) and the (x, y, z) of the 2D crystal, where the z and Z axis coincide. The laser beam is propagating along Z, is focused on the sample at normal incidence, and is linearly polarized along the X-Y plane, at angle φ with respect to X laboratory axis. Using a rotating half-waveplate, we rotate φ and detect the SHG produced by the 2D crystal as function of φ, performing P-SHG imaging. The x axis is along the armchair crystallographic direction at angle θ from X. In laboratory coordinates, the laser excitation electric field after passing the half-waveplate can be expressed as the Jones vector $\begin{pmatrix} E_0 \cos\varphi \\ E_0 \sin\varphi \end{pmatrix}$, where $E_0$ is the amplitude of the electric field, and we have assumed $E_Z^\omega = 0$, considering the excitation field polarized along the sample plane. This expression can be transformed in crystal coordinates by multiplying with the rotation matrix $\begin{pmatrix} \cos\theta & \sin\theta \\ -\sin\theta & \cos\theta \end{pmatrix}$, giving $E^\omega = \begin{pmatrix} E_x^\omega \\ E_y^\omega \end{pmatrix} = \begin{pmatrix} E_0 \cos(\varphi - \theta) \\ E_0 \sin(\varphi - \theta) \end{pmatrix}$.

Monolayer TMDs belong to the D$_{3h}$ point group and therefore their $\chi^{(2)}$ tensor exhibits four nonzero elements, namely, $\chi_{xxx}^{(2)} = -\chi_{xyy}^{(2)} = -\chi_{yyx}^{(2)} = -\chi_{yxy}^{(2)}$, where x, y, z denote the crystalline coordinates [11,13,15]. Therefore, the non-linear theoretical model governing the SHG process can be expressed in matrix form as[10,13,15]:

$$\begin{pmatrix} P_x^{2\omega} \\ P_y^{2\omega} \\ P_z^{2\omega} \end{pmatrix} = \varepsilon_0 \chi_{xxx}^{(2)} \begin{pmatrix} 1 & -1 & 0 & 0 & 0 & 0 \\ 0 & 0 & 0 & 0 & 0 & -1 \\ 0 & 0 & 0 & 0 & 0 & 0 \end{pmatrix} \begin{pmatrix} E_x^\omega E_x^\omega \\ E_y^\omega E_y^\omega \\ E_z^\omega E_z^\omega \\ 2 E_y^\omega E_z^\omega \\ 2 E_x^\omega E_z^\omega \\ 2 E_x^\omega E_y^\omega \end{pmatrix} \quad (1)$$

where $E^\omega$ represents the laser electric field, $P^{2\omega}$ represents the induced SHG polarization and ε$_0$ is the permittivity of free space. By substituting the expression of $E^\omega$ into Equation 1, and then rotate back to laboratory coordinates XY, we obtain $\begin{pmatrix} P_X^{2\omega} \\ P_Y^{2\omega} \end{pmatrix} \sim \varepsilon_0 \chi_{xxx}^{(2)} \begin{pmatrix} \cos(3\theta - 2\varphi) \\ \sin(3\theta - 2\varphi) \end{pmatrix}$. Before the detector, we have used a linear polarizer in an angle ζ with X-axis to select a specific component of the SHG field. In order to account for the effect of this polarizer, we multiply with the Jones matrix $\begin{pmatrix} cos^2\zeta & sin\zeta cos\zeta \\ sin\zeta cos\zeta & sin^2\zeta \end{pmatrix}$. Here we have set $\zeta = 0$, measuring $P_X^{2\omega} = P_{//}^{2\omega}$, whose intensity $I_X^{2\omega} = I_{//}^{2\omega} = |P_{//}^{2\omega}|^2$ is calculated as [13,15]:

$$I_{//}^{2\omega} = A\,cos^2(3\theta - 2\varphi) + C, \qquad (2)$$

where $A = \varepsilon_0^2 \left(\chi_{xxx}^{(2)}\right)^2 E_0^4$ is a multiplication factor and $C$ is a constant that accounts for experimental errors (e.g., imperfections of the optical components used in the experiments).

By using the galvanometric mirrors, we raster-scan a large area and we record SHG images of $I_{//}^{2\omega}$, for different values of the linear polarization angle φ of the laser field, with φ ϵ [0⁰, 360⁰] with step 2⁰. We then use Equation 2, to fit pixel-by-pixel the experimental data and calculate the spatial distribution of the armchair direction θ. Each image consists of 500x500 measurements (pixels). The radial resolution offered by the 40x 1.3NA objective is ~483 nm. For the data analysis, the MATLAB programming language (The Mathworks, Inc) was used.

**Data availability**

All data that support the findings of this study are included within the article (and any supplementary files).

**Acknowledgements**

G.Kour, L.M., and G.Kio., acknowledge funding by the Hellenic Foundation for Research and Innovation (H.F.R.I.) under the 'First Call for H.F.R.I. Research Projects to support Faculty members and Researchers and the procurement of high-cost research equipment grant' project No: HFRI-FM17-3034. A.M., J.P., and K.P. acknowledge support by the project SPIVAST funded by the Foundation for Research and Technology Hellas. S.P. and E.S. acknowledge financial support by the European Union's Horizon 2020 research and innovation program through the project NEP, EU Infrastructure, GA 101007417 –INFRAIA-03-2020. E.S. and G.Kio. acknowledge financial support by the EU-funded DYNASTY project, ID: 101079179, under the Horizon Europe framework programme.

**Author contributions**

S.P., E.S., and G. Kio. designed and supervised this work. G. Kou prepared and characterized the samples. G. Kou., S.P., and G.M.M. performed the PSHG experiments and data analysis. L.M. performed theoretical calculations and data analysis. J.A.C prepared the pre-patterned substrates. M.T., and A.B.D performed the AFM experiments. A.M., J.P., and K.P. performed the Raman measurements and did the corresponding analysis. G. Kou., S.P., G.M.M., L.M., E.S., and G. Kio., wrote the paper. All authors viewed and commented on the manuscript.

**Competing interests**

The authors declare no competing interests.

**Additional information**

**Supplementary Information:** The online version contains supplemental material

**Correspondence** and requests for materials should be addressed to E.S. or G.Kio.

# Supplementary Information

# Strain distribution in WS$_2$ monolayers detected through Polarization-resolved Second Harmonic Generation

George Kourmoulakis[1,2], Sotiris Psilodimitrakopoulos[1]*, George Miltos Maragkakis[1,3], Leonidas Mouchliadis[1], Antonios Michail[4,5], Joseph A Christodoulides[6], Manoj Tripathi[7], Alan B Dalton[7], John Parthenios[5], Konstantinos Papagelis[5,8], Emmanuel Stratakis[1,3]*, and George Kioseoglou[1,2]*

[1] Institute of Electronic Structure and Laser, Foundation for Research and Technology - Hellas, Heraklion, 71110, Crete, Greece
[2] Department of Materials Science and Technology, University of Crete, Heraklion, 70013 Crete, Greece
[3] Department of Physics, University of Crete, Heraklion Crete 70013, Greece
[4] Department of Physics, University of Patras, Patras, 26504, Greece
[5] FORTH/ICE-HT, Stadiou str Platani, Patras 26504 Greece
[6] Naval Research Laboratory, 4555 Overlook Ave SW, Washington, DC 20375-5320, U.S.A
[7] Department of Physics and Astronomy, University of Sussex, Brighton BN1 9RH, United Kingdom
[8] School of Physics, Department of Solid-State Physics, Aristotle University of Thessaloniki, Thessaloniki,54124, Greece

* Authors to whom any correspondence should be addressed.

Email: sopsilo@iesl.forth.gr, stratak@iesl.forth.gr, and gnk@materials.uoc.gr

## Room Temperature Raman characterization of monolayer WS₂

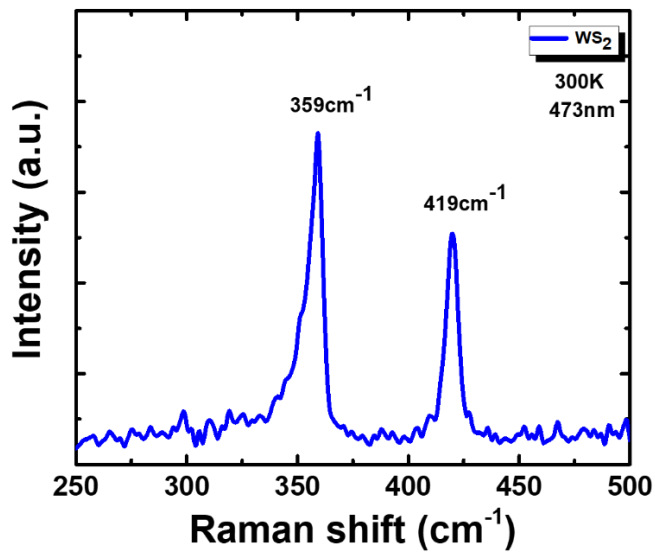

**Figure S1:** The energy difference of the prominent vibrational modes is 60cm$^{-1}$ which proves the existence of monolayer WS$_2$.

# Room temperature photoluminescence characterization for the regions of interest

Figure S2 presents a comparison of photoluminescence (PL) emission originating from both suspended and strained regions within the same monolayer of $WS_2$. The impact of $Si/SiO_2$ on the 2D materials is well known, introducing disorder through factors such as local strain resulting from the amorphous nature of the oxide layer, carrier doping, and impurities. In the context of suspended $WS_2$, the absence of a substrate contributes to an enhanced PL emission. In contrast, strained areas of $WS_2$ exhibit a noticeable suppression in PL intensity. The influence of substrate disorder is evident here, as the monolayer reaches the bottom of the cylindrical well. Additionally, strain-induced bandgap narrowing may account for the observed redshift in the PL emission energy in this case.

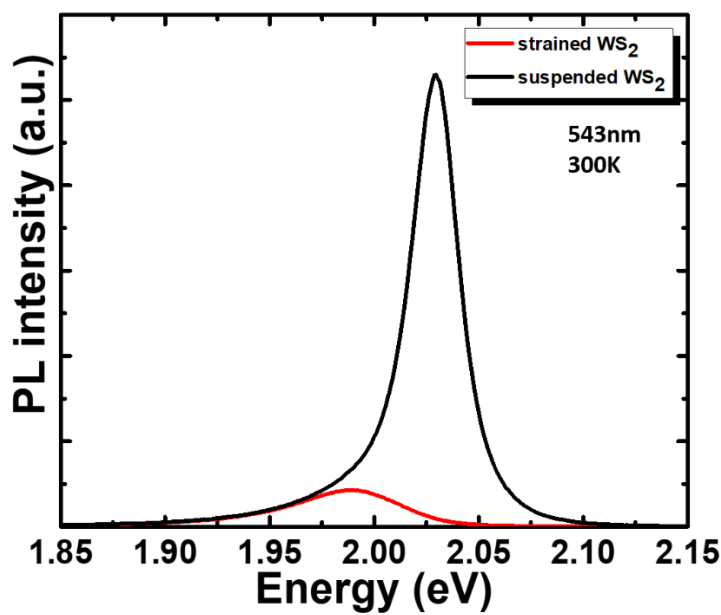

**Figure S2:** Suspended areas (black line) present a 10-fold enhancement in contrast to strained areas (red line)